\documentclass[12pt]{iopart}
\usepackage{epsfig}
\usepackage{graphicx}
\usepackage{lscape}
\usepackage{longtable}
\usepackage{deluxetable}
\usepackage{CJK}

\usepackage{iopams}  
\begin{document}

\title[Octans-Near Association and Castor Moving Group]{Young Stars Near Earth: The 
Octans-Near Association and Castor Moving Group}

\author{B. Zuckerman$^1$, Laura Vican$^1$, Inseok Song$^2$, and Adam 
Schneider$^3$ }

\address{$^1$Department of Physics and Astronomy, University of California, 
Los Angeles, CA 90095, USA}
\address{$^2$Department of Physics and Astronomy, University of Georgia, 
Athens, GA 30602-2451, USA}
\address{$^3$Department of Physics and Astronomy, University of Toledo, 
Toledo, OH 43606, USA}
\eads{\mailto{ben@astro.ucla.edu}, \mailto{lvican@ucla.edu}, \mailto{song@uga.edu}, \mailto{Adam.Schneider@Utoledo.edu}}
\begin{abstract}
All cataloged stellar moving groups and associations with ages $\leq$100 
Myr and within 100 pc of Earth have Galactic space motions (UVW) situated 
in a "good box" with dimensions $\sim$20 km s$^{-1}$ on a side.  Torres et 
al. defined the Octans Association as a group of 15 stars with age "20 
Myr?" and located $\sim$140 pc from Earth, but with average V space 
velocity -3.6 km s$^{-1}$ that is well outside of the good box.  We present 
a list of 14 Hipparcos star systems within 100 pc of Earth that we call 
"Octans-Near";  these systems have UVW similar to those of the much more 
distant Octans Association.  
The Octans-Near stars have apparent ages between 
about 30 and 100 Myr and their relationship to the Octans Association stars 
is unclear.  Six additional star systems have UVW similar to those of  
Octans-Near stars and likely ages $\leq$200 Myr.  These six systems include 
the late-type binary star EQ Peg -- 6.2 pc from Earth with likely age $\leq$100 
Myr and thus likely to be the nearest known pre-main sequence star system.  
The UVW of stars in a previously proposed $\sim$200 Myr old Castor moving 
group are not too dissimilar from the UVW of Octans-Near stars.  However, 
stars in the Castor group -- if it exists at all -- are mostly 
substantially older than 200 Myr and thus generally can readily be 
distinguished from the much younger Octans-Near stars.
\end{abstract}
\maketitle

\section{Introduction}

Beginning in the late 1990s astronomers have identified a handful of 
moving groups within 100 pc of Earth with ages $\leq$100 Myr (for 
summaries see Zuckerman \& Song 2004, Torres et al 2008, and Zuckerman et 
al 2011).  Two additional nearby moving groups -- Castor (Montes et al 
2001; Caballero 2010) and Carina-Near (Zuckerman et al 2006) -- are 
somewhat older with suggested ages $\sim$200 Myr, although, as discussed 
in Section 4.2, the proposed Castor stars are actually considerably older 
than 200 Myr.   Youthful, nearby stars are of interest as the most advantageous examples for 
study of early stellar evolution and direct imaging of self-luminous planets.  
For completeness sake, we note the presence of some older stellar associations within
100 pc of Earth.  The age of the Hercules-Lyra association is recently 
given as 257 Myr (from gyrochronology, Eisenbeiss et al 2013).  The Ursa 
Majoris moving group has a likely age somewhere in the range 400-500 Myr 
(King et al 2003; Zuckerman et al 2006), while Hyades stream stars are 
generally accorded the age of the Hyades cluster ($\sim$600 Myr).

Inspection of Table 1 in Torres et al (2008) reveals that the mean V 
component of Galactic space velocity (UVW) for known moving groups within 
$\sim$100 pc of Earth and with ages $<$100 Myr always lies between -14 and 
-28 km s$^{-1}$.  A similar picture can be seen in Figure 6 of Zuckerman 
\& Song (2004).

While investigating a mostly unpublished list of young stars 
compiled by two of us (IS and BZ) and Mike Bessell, we noted that HIP 
17338 possesses an infrared excess in the WISE catalog 
(Wide-Field Infrared Survey Explorer; Wright et al. 2010) and is obviously 
very young (see Appendix A1), yet has a V velocity well outside the range 
-14 to -28 km s$^{-1}$.  Additional study revealed that the UVW of HIP 
17338 is identical to that of the 15-member Octans Association defined by Torres et 
al (2008).  However, given its large mean distance from Earth (141 pc),
deep-southern sky-plane location, and sparseness, there is 
little reason to postulate a direct association between the Octans 
Association and HIP 17338 (which is located only 50 pc from Earth).  
None of the 15 Octans Association stars are in the Hipparcos catalog.

We therefore undertook a search for Hipparcos stars within 100 pc of Earth 
that might have ages and UVW similar to that of HIP 
17338.  The results of this search are presented in Section 3 and Tables 1 
and 2.  Because most of these stars are probably older than the "20 Myr?" 
Torres et al (2008) attribute to their Octans Association, and because of 
the very different sky-plane and parallax distribution of the Torres et al 
stars and the Table 1 stars, we dub the latter stars the "Octans-Near 
Association".

The UVW of the Castor moving group, as considered by Montes et al (2001), 
is not too dissimilar from that of Octans-Near.  
However, as discussed in Section 4.2, if the Castor group 
exists at all, then it is substantially older than Octans-Near.  Thus, 
distinguishing whether a given star might belong to Octans-Near or to 
Castor typically is not difficult.

\section{Search Method and Observations}

The search criteria were simple.  For Hipparcos stars with known radial 
velocities and within 100 pc of Earth we calculated UVW velocities and 
compared these to those of the Octans Association (which are given in the 
notes to Table 1).  If all three components of velocity for a given star 
agreed within about twice the sum of the velocity errors for the star and 
the velocity dispersion of the Octans Association, then we further 
investigated the star to determine its age via methods described in 
Section 3.

We used the Hamilton Echelle and Kast grating spectrometers (Vogt 1987) on 
the 3-m Shane telescope at Lick Observatory to obtain spectra of many of 
the stars discussed in Section 3 and Table 1.  A primary motivation was to 
measure the strength of the lithium absorption line at 6708 \AA\ which is 
often a good youth diagnostic for stars of spectral type mid-F and later 
(e.g., Zuckerman \& Song 2004).  The spectrometer parameters and observing 
dates are given in Table 3.  All data were reduced using standard IRAF 
packages. Equivalent widths (EW) were measured by fitting a Voigt profile 
in IRAF to spectral features.  With the Hamilton we could measure an EW of 
$\sim$10 m$\AA$ when a signal-to-noise of $\sim$50 was reached.  Where the 
words "this work" appear in column 5 of Table 2, then the quoted lithium 
EW are from Hamilton spectra; Kast spectra EW are always in reasonable 
agreement.  A representive spectrum from each spectrometer appears in 
Figure 1.

All primary stars in Tables 1 and 2 for which lithium spectra were 
obtained are bright and the signal-to-noise was always large.  All lithium 
EW were measured independently from the spectra by either two or three 
persons; based on the dispersion in these measurements we estimate an 
internal EW error of $\pm$10 m\AA.  External errors are usually comparable 
as may be seen by comparing our lithium EW with those measured by others 
for some stars listed in Appendix A; the only star where the total dispersion 
in measured EW is larger than 20 m\AA\ is HIP 110778.

We calculated the vsini of stars in our sample by measuring the FWHM of 
absorption features near 6430$\AA$, following the procedure outlined in 
Strassmeier et al. (1990). Typical uncertainties are 2-3 km s$^{-1}$. 
These velocities are listed in column 3 of Table 2, while the literature 
values (largely taken from Glebocki \& Gnacincki 2005) are listed in 
column 2.

\section{Results}

We calculated the UVW in Table 1 with radial velocities from Gontcharov 
(2006) and occasionally other sources, and parallaxes and proper motions 
from Hipparcos (van Leeuwen 2007).  As noted in Section 2, the UVW of 
stars tested for membership in Octans-Near were initially taken to 
(nearly) match those of the Octans Association proposed by Torres et al 
(2008).  The value of the means of UVW, for the Octans Association and for 
Octans-Near stars listed in Section A of Table 1, are given in the notes 
to Table 1.  Because the binary star HIP 813 is obviously youthful (see 
Appendix A1), it is included in Section A even though its V velocity 
appears to be somewhat discrepant from those of the Octans Association and 
of Octans-Near.  In addition to the usually derived error in V given in 
Table 1, for HIP 813 there is an additional uncertainty of $\leq$0.5 km 
s$^{-1}$ due to the orbital motion of the binary.  In the event that HIP 
813 is not included as an Octans-Near star, then the mean V velocity of 
the other 13 systems is -4.0$\pm$1.5 km s$^{-1}$.

When HIP 813 is included, the dispersion in each component of UVW for the 
proposed Octans-Near stars in Section A of Table 1 is, on average, about 
1.5 times those of associations listed in Tables 1 and 2 of Torres et al 
(2008).  However, these associations are mostly more compact, or 
closer to Earth, or younger than stars in Table 1A.  The 
lower three panels in Figure 2 display the dispersion in UVW for stars 
listed in Sections A and B of Table 1, as well as some stars from the 
postulated Castor moving group (see Section 4.2).  The upper panels in Figure 
2 display the distribution in the sky of the same stars in XYZ 
coordinates, where X, Y and Z are directions that correspond to those of 
UVW (given in the notes to Table 1), and with their origin at the Sun.

Evidently, the proposed Octans-Near stars have a substantial spread in XYZ;
this spread is similar to those of some associations listed in Table 2 and
displayed in figures in Torres et al (2008), including the Octans Association itself.
What distinguishes Octans-Near from most of these other associations (although not from the 
Octans Association itself) is the relatively low space density (stars per cubic pc) 
of Octans-Near.   As noted in Section 2, we considered only Hipparcos stars with known radial
velocities.  Therefore, this apparent low space 
density may be, at least in part, a selection effect. That is, the stars listed in Table 1A are clearly 
"top-heavy" in mass, being dominated by A-type stars.  This distribution is not to be expected   
in any normal star formation scenario and, therefore, many K- and M-type
Octans-Near stars may be discovered when parallaxes of additional low mass 
stars are measured.

For stars later than early-F type, age estimates in Table 1 are based 
primarily on lithium 6708 \AA\ equivalent width (EW) and to a lesser
extent on activity indicators.
Figure 3 displays Li 6708 \AA\ EW for F-G- and K-type stars listed in 
Table 1, Sections A and B, along with Octans Association stars (da Silva 
et al 2009) and Pleiades cluster members (Soderblom et al 1993).  It 
appears that four of the Octans-Near systems might be as young as the 
proposed 20 Myr age of the Octans Association (Torres et al 2008); this 
includes CD-58 860 which is the open diamond symbol.  All other plotted 
Octans-Near stars from Table 1, Sections A and B, appear to be as old or 
older than the Pleiades and the same may be said for one or two Octans 
Association stars.

For late-B and A-type stars in Table 1A, we use placement on a 
color-magnitude diagram (Zuckerman 2001, Vigan et al 2012, Zuckerman \& 
Song 2012, Nielsen et al 2013).  Beginning with Jura et al (1998) and 
Lowrance et al (2000) a low placement has been interpreted as indication 
of a youthful A-type star and, as may be seen in Figure 4, the A-type 
stars in Table 1A all plot near the locus of $\sim$100 Myr old Pleiades 
members.  However, Nielsen et al (2013) demonstrate that such a placement 
is only a necessary, but not a sufficient, condition for youth.  Of the 
seven B- and A-type stars in Table 1A, six have either a young late-type 
companion or excess infrared emission, all of which point to a youthful 
age consistent with their placement on the color-magnitude diagram.  Details may 
be found in Section 4.1 and Appendix A1.  However, neither placement on a 
color-magnitude diagram nor the presence of excess infrared emission 
(i.e., a debris disk) can do more than indicate that any individual 
early-type star is likely to be $\sim$100 Myr old.

To better understand the applicability of stellar magnetic activity as an 
age indicator for youthful stars, we compared the fractional X-ray 
luminosity (L$_{x}$/L$_{bol}$) of our candidate Octans members with that 
of nearby young moving groups (Figure 5). The X-ray luminosity traces the 
coronal magnetic activity, which typically decreases as a function of age 
and stellar mass (Zuckerman \& Song 2004, Mamajek \& Hillenbrand 2008).  
We used ROSAT Bright and Faint source catalogs to obtain X-ray count rates 
and then the relation given in Schimitt et al. (1995) to convert to X-ray 
fluxes. Visual magnitudes, B-V colors, and the bolometric correction 
values given in Kenyon and Hartman (1995) were used to calculate stellar 
bolometric fluxes. By plotting B-V (which traces stellar mass) against 
L$_{x}$/L$_{bol}$, we see that young stars in nearby moving groups 
($\beta$ Pic - 12 Myr; Tuc-Hor- 30 Myr; AB Dor - 70 Myr) 
lie along one path while older stars in the 
Hyades Cluster ($\sim$600 Myr) follow another. The X-ray ages listed in 
Table 1 were estimated from Figure 5.

We considered using the X-ray - age relations from Mamajek \& Hillenbrand 
(2008; see their Equation A3). Their equation for an X-ray age is based on 
an empirically determined relation between chromospheric activity and age 
(their Equation 3) and an observed correlation between X-ray activity and 
chromospheric activity (their Equation A2). However, the candidate 
Octans stars are already surmised (from lithium EW) to be quite young 
($\le$100 Myr) and many fall in the ``very active" 
regime (logR'$_{HK}$$>$-4.35 and log(L$_{x}$/L$_{bol}$)$>$-4.0) where the 
correlation between X-ray and chromospheric activity begins to break down 
(here, R'$_{HK}$ specifies Ca II H \& K emission; Table 2). We therefore 
choose to use the empirical relationship from Zuckerman \& Song (2004) to 
estimate X-ray ages (see previous paragraph).

We used the chromospheric activity relation from Mamajek \& Hillenbrand 
(their Equation 3) to estimate the activity age of stars in our sample 
with logR$^{\prime}$$_{HK}$ values available in the literature.  While we 
used these ages to provide general support for stars being older or younger than 
$\sim$100 Myr (see Vican \& Schneider 2013 for further details), we gave 
them relatively low weight in comparison with lithium-based ages.

To summarize the discussion of age diagnostics, of the methods that can be 
applied to the Octans-Near stars we regard lithium abundances, as per 
Figure 3, to give the most reliable and quantitative ages.  From this 
presentation of F-, G- and K-type stars, it is clear that most of the 
Octans Association stars proposed by Torres et al (2008) and some of the 
proposed Octans-Near stars are younger than the Pleiades, while other 
Octans-Near and Octans Association stars are as old as or somewhat older 
than the Pleiades.  For F-, G- and K-stars this young, activity ages are 
not as diagnostic.  For B- and A-type stars, while no single age dating 
method is as precise as Li abundances, a combination of low placement on a 
color-magnitude diagram and the presence of a substantial dusty debris 
disk and/or an active companion star point to youth.

A star is included in Tables 1 and 2 if each component of UVW is consistent 
with that of the association mean.  "Consistency" in this context implies 
that all 3 stellar components of UVW match those of the association mean 
within the sum of the quoted error in stellar velocity and the standard 
deviation of the dispersion among the association velocities.  However, 
because errors in UVW can be underestimated (see next paragraph), 
because a standard deviation is not an absolute limit, and because these 
tables list "proposed" association members, we have tended to err somewhat 
on the side of including a potential non-member in preference to losing a 
true member.

Given errors in measuring radial velocities, especially of early A-type stars, and, 
to a lesser degree, errors in Hipparcos parallaxes, some proposed members 
will probably ultimately turn out to be non-members.  In the present work 
and that of our earlier papers (Zuckerman et al 2011; Zuckerman \& Song 
2012) we have adopted the Gontcharov (2006) radial velocities and their 
errors while recognizing that these errors are likely to be too optimistic 
for some stars.  We adopt Hipparcos parallaxes from van Leeuwen (2007) 
while anticipating that some of these may not be correct (for example, HIP 
30675 considered by Zuckerman \& Song 2012).

\section{Discussion}

In the spirit of Zuckerman \& Song (2004), we searched for Hipparcos stars 
with common Galactic space motions within 100 pc of Earth and with age 
$\leq$100 Myr.  The result is the proposed Octans-Near Association in part 
A of Tables 1 and 2.  These stars appear to have ages between about 30 and 
100 Myr.   Notwithstanding   
their similar UVW, it is unclear how stars with such a large spread
($\sim$70 Myr) in apparent age as those listed in part A of Table 1 could
have their origins in a given molecular cloud or cloud complex. In
Appendix A we consider individual stars in Table 1.

Of the Octans Association stars proposed by Torres et al (2008) 
only CD-58 860, at a quoted distance of 82 pc, satisfies our proximity 
requirement.  Its spectral type (G6) and 6708 \AA\ lithium EW (225 m\AA) 
suggest an age $\sim$30 Myr (the open diamond in Figure 3).  While CD-58 
860 is not a Hipparcos star, for ages in the range 30-100 Myr, a 
photometric distance $\sim$90-100 pc can be derived from Figure 2 in 
Zuckerman \& Song (2004).  The best match with the UVW of Octans-Near (and 
the Octans Association) is found for a distance $\sim$90 pc.  The distance
from Earth and sky-plane position (4h11m -58d) of CD-58 860 are not far
from some stars listed in Table 1A.  Thus, CD-58 860 may reasonably 
be considered an Octans-Near member. 

\subsection{Infrared and X-ray Emission}

As noted in Appendix A1, three of the stars in part A of Table 1 have 
previously reported excess infrared emission above the photosphere (Su et 
al 2006; Rhee et al 2007).  We searched all stars in Table 1 for excess IR 
emission in one or more of the four WISE bands.  Spectral energy 
distributions (SEDs) were fit to the optical and IR flux densities as 
described in Vican \& Schneider (2013).  SEDs were created with an 
automated fitting technique that fits the stellar radius and temperature to 
the Hauschildt et al. (1999) photosphere models. A simple 
single-temperature blackbody was then fit to the dust excess. The resulting 
SEDs are found in Figure 6 for the five Table 1A stars with definite 
excesses, and in Figure 7 for the two stars with possible excesses (see 
discussion below). All WISE data products were checked to make sure that 
the apparent 22 $\mu$m (W4) excess was not due to a background galaxy or 
cirrus contamination.

Because the SED fitting procedure suggested the possibility of small (but 
individually not statistically significant) W4 excesses for many of the 
Octans stars, we carried out an additional excess check that utilized only 
the W3 (11.6 $\mu$m) minus W4 magnitude. This check involved Octans stars 
and, also, stars measured with the Spitzer Space Telescope, as described in 
the paragraphs that follow. Based on the WISE calibration procedure, for 
late B- to early K-type stars as appear in Table 1, we anticipated that, 
for the ensemble of stars without excess emission in the W4 band, the W3 
minus W4 color (in Vega magnitudes) would be zero.  We defined an excess 
sigma ($\sigma$) for a given star as (W3-W4)/e, where "e" equals the mean 
W4 error given in the WISE catalog.

As noted two paragraphs below, 7 stars in part A of Table 1 individually 
show no obvious evidence for excess emission in W4. Nonetheless there is a 
positive bias among these 7, with $\sigma$ ranging between 0 and 2.5 for 
the individual stars and with a mean of 1.55 for the 7 stars.  Similarly, 
for the 13 stars in Table 1 part B plus part C, the mean $\sigma$ is 1.64 
(the 13 stars include HIP 66704A \& B and GJ 4185A \& B).

To check whether this apparent red W3-W4 color might indicate a small 
average IR excess in Table 1 stars or, alternatively, a small calibration 
error in the WISE 3 or WISE 4 zero-point flux density, we examined the 
W3-W4 color of 19 stars without 24 $\mu$m, and usually also without 70 
$\mu$m, excess emission in a 
Spitzer-measured sample discussed in Zuckerman et al (2011).  These 19 
stars are likely members of the AB Dor, Tuc/Hor, and Argus young moving 
groups which have ages similar to those of the 14 stars in part A of Table 
1.  The mean $\sigma$ for the 19 Spitzer-measured stars is 1.55 with a 
standard deviation of 0.92 and a mean error of 0.22.  This comparison 
suggests that there is a few $\%$ calibration error in the relative W3 and 
W4 zero-points and that an apparent excess in the W4 band smaller than 
about 5 $\sigma$, (with $\sigma$ as defined above), is not significant. 
    
Of the 14 primary stars in part-A of Tables 1 and 2, five -- HIP 
10670, 17338, 19990, 22192, 36624 -- definitely have 
excess infrared emission indicative of a dusty debris disk (Figure 6).  All 
five stars, with the possible exception of 19990, 
have excess emission in the 22 $\mu$m W4 band of WISE.
Of the other 9 stars, two -- HIP 813, 115527 -- might have excess emission 
in the W4 band (Figure 7).  If, say, one of these two possible excess 
stars, actually has an excess, then 43\% of stars in part-A have excess IR 
emission.  Such a high percentage is consistent with ages $\leq$100 Myr 
(e.g. Table 11 in Zuckerman et al 2011 which focuses on stars in clusters 
and moving groups).

If only the seven A- and B-type stars in part A are considered, then four 
have definite IR excesses and one other (HIP 813, which surely is young, 
see Appendix A1) might also; this is again a very high percentage of 
detectable dusty debris disks, suggestive of youth.  Su et al (2006) 
consider excess 24 $\mu$m emission from A-type stars in clusters and in the 
field; their Figure 4 depicts the ratio (R$_{24}$) of the observed 24 
$\mu$m flux 
density to that expected for the photospheric emission.  For the five A and 
B stars in Table 1A listed in the previous paragraph that have definite or 
possible excess emission at W4, the mean R$_{24}$ is $\sim$1.4.  As may be 
seen from the Su et al Figure 4, this (large) value of R$_{24}$ is 
primarily characteristic of stars with ages $\leq$100 Myr.

HIP 16095 is one of the two A- and B-type stars in Table 1A with neither 
excess infrared emission nor a resolved, young, late-type companion.  
However, HIP 16095 is an X-ray source (see Appendix A1) and very likely 
has an active, unresolved, youthful companion. Thus, six of the seven A- 
and B-type stars in Table 1A possess a youth indicator in addition to 
their low placement on the color-magnitude diagram.

\subsection{The Castor Moving Group}

Of the moving groups and associations listed in Zuckerman \& Song (2004) 
and Torres et al (2008) none have UVW similar to that of Octans-Near.  
However, a Castor moving group has been considered by various authors 
(e.g., Anosova \& Orlov 1991; Barrado y Navascues 1998; Montes et al 2001; 
Caballero 2010).  The number of proposed candidate Castor group members in 
these various papers range from 10 separate systems (Anosova \& Orlov 1991)  
to $\sim$50 systems (Caballero 2010).  Of 26 systems considered by Barrado 
y Navascues (1998), he concludes that only 16 are physically associated and 
he expresses doubts as to whether a Castor moving group exists.  
Notwithstanding this uncertainty, he proposes a UVW for the group of -10.7, 
-8.0, -9.7 (3.5, 2.4, 3.0) km s$^{-1}$ which was subsequently adopted by 
Montes et al (2001) and by Caballero (2010).  Anosova \& Orlov (1991) 
include Fomalhaut and its companion GJ 879 as members of their proposed 
Castor moving group.  Barrado y Navascues includes Fomalhaut, and also adds 
Vega; he suggests ages of 200$\pm$100 Myr for these two famous stars.  
This age was then adopted by Montes et al (2001) and by Caballero (2010).  
While Montes et al do not explicitly list A-type members of their proposed 
moving groups, Fomalhaut would be included by virtue of their inclusion of 
TW PsA which is "unequivocally a physical stellar companion to Fomalhaut" 
according to Mamajek (2012). Caballero explicitly includes Vega, Fomalhaut 
and TW PsA as candidate Castor moving group members.

We identify some problems with a Castor moving group as discussed by the 
above authors.  While the proposed age has been 200 Myr, the best recent 
age estimates are 440$\pm$40 and 700+150/-75 
Myr for Fomalhaut and Vega, respectively (Mamajek 2012; Monnier et al 
2012).  To further check the age of the proposed Castor group we used the 
Hamilton echelle spectrometer to measure the lithium 6708 \AA\ line in PR 
Vir, GY Boo, MS Dra, and BD+21 3245.  These K0 to K3 type stars are Castor 
moving group candidates according to Caballero (2010).  In all cases the 
lithiium line is undetected with upper limit of order 10 m\AA.  For early 
K-type stars this would correspond to stars no younger than the UMa group 
with an age of 400-500 Myr (see figures in Zuckerman \& Song 2004 and 
Zuckerman et al 2006).

For an age of 400 Myr a difference in Galactic V velocity of even 3 km 
s$^{-1}$ would result in a spread in location in the Galactic plane of a 
kpc.  Yet differences in V among many of the proposed Castor group stars 
are greater than 3 km s$^{-1}$ (e.g., Table 6 in Montes et al 2001).  For 
example, for Vega and Fomalhaut we calculate respective UVW of -18.4,
-12.4, -9.7 (0.1, 0.2, 0.1) km s$^{-1}$ and -5.8, -8.3, -11.0 (0.2, 0.1, 
0.5) km s$^{-1}$, yet these two stars are within 15 pc of each other.  As may
be seen in the lower panels of Figure 2, the UVW dispersion of Castor stars is 
substantial.

\section{Conclusions}

We have identified 17 late-B through early-M type stars in 14 separate 
systems within 100 pc of Earth with similar Galactic space velocities UVW 
and ages in the range 30-100 Myr.  Because these UVW are similar to those 
of a previously defined Octans Association at much larger distance from 
Earth (Torres et al 2008), we dub these newly identified youthful stars 
the "Octans-Near" Association.  Of the 15 Octans Association stars 
proposed by Torres et al, only CD-58 860 also qualifies as an Octans-Near 
member.

If a moving group is envisioned as a set of stars that were born more or 
less together in a given molecular cloud or cloud complex, then the likely 
large spread in ages ($\sim$70 Myr) and sky position of the proposed 
Octans-Near stars is troublesome.  Octans-Near has no obvious "nucleus" 
such as characterizes the Tucana Association (Zuckerman \& Webb 2000), the 
AB Dor moving group (Zuckerman et al 2004; Barenfield et al 2013), and the 
Carina-Near moving group (Zuckerman et al 2006). The proposed Octans-Near 
stars also have a somewhat (factor of $\sim$1.5) larger dispersion in 
Galactic space velocities (UVW) than do the associations listed in Torres 
et al (2008).  In the 1960s, Eggen introduced the idea of a Local 
Association of youthful stars (e.g., Eggen 1965 and citations of earlier papers by 
Eggen). Eggen's extensive Local Association, also referred to as the 
Pleiades moving group, was envisioned to include both field stars and 
stars in some well known clusters and, in this sense, differed greatly 
from our proposed, sparse, Octans-Near Association.

The referee has suggested that, given the above issues, the stars in Table 
1 might best be called the "Octans-Near complex".  Ultimate resolution of 
whether the Octans-Near stars actually form a true moving group may come 
from future accurate parallax and radial velocity measurements.  
Notwithstanding this uncertainty, the Octans-Near stars are sufficiently 
young and close to Earth that they should be good targets for direct 
planet-imaging searches with extreme adaptive optics systems on large 
telescopes.

Whatever the outcome of future observations, there is little doubt that a 
significant number of nearby stars with ages $\leq$100 Myr exist with UVW 
near -13, -4, -11 km s$^{-1}$ which is distinctly different from those of 
any moving group within 100 pc of Earth considered by Zuckerman \& Song 
(2004) and Torres et al (2008).  Perhaps as many as 50\% of the 14 
Octans-Near primary stars display excess infrared emission above the 
stellar photosphere in the WISE4 (22 $\mu$m) band.  We examined the WISE3 
(11.6 $\mu$m) and W4 band magnitudes of Octans-Near stars and also stars 
previously measured with the Spitzer Space Telescope that lack 24 $\mu$m 
and, usually, 70 $\mu$m excess emission.  From this comparison we deduce 
that there is a few percent relative zero-point calibration offset between 
the W3 and W4 bands.  This offset was taken into account when we evaluated 
whether or not a given Octans-Near star had excess emission in the W4 
band.

The possibility of a moving group associated with the bright, multiple, 
star Castor has been considered in the literature for over two decades.  
Based in part on recent age estimates for some proposed Castor moving 
group members and on our own lithium abundance measurements of some other 
proposed members, we conclude that there exists no convincing evidence for 
a Castor moving group.  This putative group (see Caballero 2010 for a 
recent accounting of proposed group members) suffers from two important 
flaws.  First is a wide spread in ages amounting to hundreds of Myr among 
proposed members.  Second is a spread of many km s$^{-1}$ in the V 
component of Galactic motion.  For a typical Castor stellar age of 
$\sim$400 Myr, a spread of even 3 km s$^{-1}$ would disperse coevally born 
stars over a Galactic arc of length of order a kpc.

We thank Michael Bessell and David Rodriguez for obtaining spectra of HIP 105044 and the 
referee for a timely and helpful report that substantially improved the 
paper.  This research was funded in part by NASA grants to UCLA and the 
University of Georgia and by a National Science Foundation pre-doctoral 
fellowship to Laura Vican.

\appendix

\section{Individual Stars}

Here we consider the stars in Tables 1 and 2.  Assignment of a given star 
into parts A, B or C of these tables is based primarily on how the 
estimated age compares with 100 Myr (or less) and the similarity between 
the UVW of the star and that of the mean UVW for the proposed Octans-Near 
Association; this mean UVW (see footnote to Table 1) is essentially 
identical to the UVW of the much more distant Octans Association defined 
by Torres et al (2008).  As we have noted, the spatial location and 
estimated age of many Octans-Near stars differ substantially from Octans 
Association stars.  Therefore, the two groups need not have identical UVW.  
Nonetheless, most of the stars in parts B and C of Tables 1 and 2 appear 
to be both older and have UVW that differ noticeably in at least one 
component of velocity from that of Octans-Near.  That is, for stars within 
100 pc of Earth, the further one deviates from the UVW of the Octans 
Association, the older do the stars appear to be.

\subsection{Probable Octans-Near Stars}

HIP 813: A B9 + G6 $\sim$8" separation 
binary star noted by Lindroos (1985).  The primary lies on the $\sim$100 
Myr locus of Pleiades age stars (Figure 5 in Zuckerman 2001).  The colors 
and Li EW of the secondary have been investigated by Pallavicini et al 
(1992) and Martin et al (1992) who both classify it as a G5V 
star.  Measurement of the Ca II K-line in emission (EW = 2.2 \AA), a 
filled H$\alpha$ line, and a 6708 \AA\ lithium EW $\sim$290 m\AA\
(Pallavicini et al 1992; Martin et al 1992), are consistent with an age 
$\sim$30 Myr for the secondary (Figure 3).  Gerbaldi et al (2001)
place the secondary on 30 Myr isochrones. 
The effective temperature of the secondary is $\sim$5300 K (Martin et al 
1992;  Gerbaldi et al 2001).  
Pecaut \& Mamajek (2013) consider the colors and temperatures of pre-main 
sequence stars.  According to their Table 6, a 30 Myr old G6 star gives a 
reasonable match to the 
effective temperature and to the 2MASS J-K magnitude and implies a B-V of 
0.74.  However, with a V magnitude of 10.4 (Pallavicini et al
1992 and Martin et al 1992) and 2MASS K-magnitude of 8.18, 
then V-K = 2.22 -- this corresponds 
to a young K0 star according to Pecaut \& Mamajek (2013).  V-K can be 
reconciled with the other data if there is $\sim$0.3 magnitudes of 
extinction at V.  B-V = 0.74 and a ROSAT X-ray luminosity, 
log(L$_x$/L$_{bol}$), = 
-3.2 together imply a star of age $\leq$100 Myr (Zuckerman \& Song 
2004), in agreement with the age of the system suggested by Montesinos et 
al (2009).  There is possible excess 
emission above the photosphere in the 22 $\mu$m WISE band (Figure 7).

HIP 10670:  A dusty debris disk is present (Rhee et al 2007).  There is 
excess emission above the photosphere in the 22 $\mu$m WISE band (Figure 6).  
An age of 100 Myr is consistent with placement on an A-star color magnitude 
diagram (Figure 4).

HIP 14007:  Cutispoto et al (2002) report this to be an SB1 composed of G3 
and K5 stars with M$_V$ = 5.2 and 8.0, respectively.  The lithium EW 
listed in our Table 2 is an average of those given in Cutispoto et al (145 
m\AA) and in Torres et al (2006; 120 m\AA).  The EW from Cutispoto et al 
(2002) is incorrectly listed as 14.5 m\AA\ in the Vizier on-line data.  
The Li EW is appropriate for a G3 star of about Pleiades age (Figure 3).

HIP 16095: This star sits very low on the A-star color magnitude diagram 
(Figure 4).  A0 type stars 
are not intrinsic X-ray sources, yet HIP 16095 was detected in the ROSAT 
all-sky survey (Schroeder \& Schmitt 2007) strongly implying the existence 
of a late-type secondary.  At a distance of 88 pc, the X-ray luminosity is 
2 x 10$^{29}$ ergs s$^{-1}$.  For a G-, K- or M-type star this X-ray 
luminosity corresponds to a star of Pleiades age or younger (see Figure 4 
in Zuckerman \& Song 2004). The likely age of HIP 16095 is $\leq$100 Myr.

HIP 17338:  The Li 6708 \AA\ EW in Table 2 is from our Hamilton spectrum; 
Torres et al (2006) give 260 m\AA\ and White et al (2007) give 243 m\AA.  
The Li EW is consistent with an age as young as 30 Myr (Figure 3).  The Ca 
II K- and 
H-line cores are in emission in our November 2012 Hamilton spectrum with 
respective EW equal to 3.0 and 2.2 \AA.  Vianna Almeida et al (2009) 
suggest this star as a possible Octans Association member.  There is 
excess emission above the photosphere in the 22 $\mu$m WISE band.

HIP 19496:  This star is headlined in SIMBAD as a member of the Pleiades 
cluster, but it has the wrong distance and proper motion to be a Pleiad.  
Rosvick et al (1992) identify the star as not a Pleiades member.  For B-V 
= 0.44, the Li EW indicates a star as young as 30 Myr (Figure 3).

HIP 19990:  A dusty debris disk is present (Su et al 2006).  An age of 100 
Myr is consistent with placement on an A-star color magnitude diagram 
(Figure 4).  Tetslaff et al (2011) give an age of 13 Myr.

HIP 22192:  A dusty debris disk is present (Su et al 2006).  There is 
excess emission above the photosphere in the 22 $\mu$m WISE band.  HIP 
22192 is a $\lambda$ Boo star that plots on
an A-star color-magnitude diagram near the 12 Myr old star $\beta$ 
Pictoris (Figure 5 in Zuckerman 2001).  As noted 
by Marois et al (2008), on such plots B-V for $\lambda$ Boo stars should be 
increased a bit to correct for low metallicity; in that case HD 22192 plots 
near the Pleiades line (see caption to Figure 4). The $\lambda$ Boo stars 
are generally thought to 
have ages up to a few 100 Myr (Gray \& Corbelly 2002).  Su et al (2006) and 
Tetzlaff et al (2011) give ages of 10 and 13 Myr, respectively.

HIP 36624: This star sits low on the A-star color magnitude diagram 
(Figure 4).  There is 
excess emission above the photosphere in the 22 $\mu$m WISE band.

HIP 73765: the Li EW is consistent with age 100 Myr (Figure 3).

HIP 97255:  This is a 10" north-south binary where the secondary, of 
spectral type about M0, is a rapid rotator (Table 2) with lithium EW $<$10 
m\AA.  Many emission lines are present in the November 2012 Hamilton 
spectrum of the secondary:  Ca II K-line (EW = 15.8 \AA), Ca II H-line (14 
\AA), H $\epsilon$ (5.3 \AA), H $\delta$ (2.4 \AA), H $\gamma$ (1.9 \AA), 
H $\beta$ (2.3 \AA), H $\alpha$ (2.3 \AA).  In the spectrum of the G0 
primary the Ca II K- and H-line cores are in emission with respective EW 
equal to 0.9 and 0.8 \AA, respectively.

HIP 102333:  An age of 100 Myr is consistent with placement on an A-star 
color magnitude diagram (Figure 4).  Chen et al (2005) give age of 50 Myr.

HIP 105044: The lithium EW in Table 2 is from a spectrum with 
resolution 7000 obtained with the Wide Field Spectrograph
(WiFeS; Dopita et al 2007) on the 2.3-m telescope at Siding Spring
Observatory (M. S, Bessell private communication 2013).  This EW 
and the X-ray luminosity are consistent with age 100 Myr.

HIP 115527:  The Hamilton and Kast spectra in the vicinity of the 6708 
\AA\ Li line are shown in Figure 1.  The Li EW and X-ray luminosity are 
consistent with age 100 Myr.  
Torres et al (2006) give Li EW = 133 m\AA. Tetzlaff et al (2011) give an age 
of 36 Myr.  There is possible excess emission above the photosphere in the 
22 $\mu$m WISE band (Figure 7).

\subsection{Possible Octans-Near Stars}

HIP 29873:  The Li EW suggests an age of 100 to a few 100 Myr.  The X-ray 
luminosity is consistent with an age of $\sim$300 Myr.  In our November 12 Hamilton 
spectrum, the deep Ca II K- and H-absorption lines have emission cores 
with respective EW equal to 0.3 and 0.13 \AA.

HIP 44526: K2 star V405 Hya. The Li EW indicates a star of likely age 
100-200 Myr.  Cabellero (2010) considers the star to be a Castor moving 
group candidate.  The Ca II K- and H-line cores are in emission in our 
November 2012 Hamilton spectrum with respective EW equal to 2.6 and 2.3 
\AA.  There is a suggestion of excess emission in the 22 $\mu$m WISE band.

HIP 66704: 17" F8 + K7 binary star.  The primary has a dusty debris disk 
(Chen et al 2005); they estimate an age of 200 Myr.  Barnes (2007) 
suggests an age 206$\pm$22 Myr.  The Li EW and X-ray luminosity are 
consistent with ages of a few 100 Myr.  The V component of velocity may be 
discordant with Octans membership.  In our November 2012 Hamilton spectrum 
of the K7 secondary the Ca II K- and H-lines are in emission with 
respective EW equal to 3.7 and 3.6 \AA.

HIP 104864:  Li EW = 85 m\AA\ (Torres et al 2006 measured 80 m\AA).  The 
X-ray luminosity is characteristic of a star $\leq$100 Myr old;  but the 
lithium age is more like a few 100 Myr (Figure 3).

GJ 4185:  The parallax and radial velocity of this previously known 26" 
binary were measured by Shkolnik et al (2012) who then calculated the UVW 
listed in Table 1 of the present paper.  With adaptive optics Shkolnik et 
al resolved the primary into two stars with delta H-magnitude of 0.37.  
Shkolnik et al construct a color-magnitude plot of absolute USNO I-mag vs I 
minus K$_s$ color that includes low-mass stars within 25 pc of Earth and 
isochrones from a 1998 paper by Baraffe et al.  Rodriguez et al (2013) 
construct a color-magnitude plot of absolute K-mag vs J minus W2 color 
(where W2 is the 4.6 $\mu$m WISE band) that includes a variety of stars, 
isochrones from the 1999 NextGen models of Hauschildt et al, and also 
empirical isochrones based on stars of known age.  Based on these two 
color-color plots we estimate an age for the binary of $\sim$150 Myr.

HIP 110778:  This is a few second of arc binary composed of two solar-type 
stars.  The proper motions in the Tycho 2, PPMXL, UCAC4 and Hipparcos 
catalogs do not agree.  In addition, a range of radial velocities appear 
in Vizier and in papers by Maldonado et al (2010) and Torres et al (2006).  
Setting the proper motion at 250, -25 mas/yr and the systemic radial 
velocity at +1 km s$^{-1}$, the UVW is -18, -7, -14 km s$^{-1}$.  We 
measure Li EWs of 122 and 143 m\AA\ for the primary and secondary 
respectively, while Torres et al (2006) measured 110 m\AA\ for both stars 
and Maldonado et al (2010) measured 83 m\AA\ (for the primary?).  A Li EW 
of 120 m\AA\ is consistent with age 100-200 Myr.

HIP 116132:  also called EQ Peg and GJ896A \& B, is a 4" binary composed 
of M3.5 and M4.5 stars.  Since this is a variable star, V-K is not well 
characterized with entries from VizieR.  M$_K$ is 6.4 and 7.3 for A and B, 
respectively.  Based on entries for V in VizieR, the minimum V-K are 5 and 
6.15 for A and B, respectively; the latter is adopted by Riedel et al 
(2011).  Even for these minimum values of V-K, the two stars plot above 
the locus of old late-type main sequence stars and near the 100 Myr 
isochrones shown in Figure 2 of Zuckerman \& Song (2004).  Also, GJ896A 
plots near an AB Dor star -- of age $\leq$100 Myr -- in Figure 3 of Riedel 
et al (2011).  The X-ray luminosity, log(L$_x$/L$_{bol}$), = -2.73; this 
is characteristic of a star of Pleiades age or younger (Figure 4 in 
Zuckerman \& Song 2004).  Therefore, HIP 116132 is likely to be $\leq$100 
Myr old, and thus, at a distance of 6.2 pc, the pre-main sequence star 
closest to Earth, even closer than AP Col (distance 8.4 pc, see discussion 
in Riedel et al 2011).  With a radial velocity of 1.5$\pm$0.4 km s$^{-1}$, 
the UVW of HIP 116132 is
-13.6, -6.0, -7.0 (0.2, 0.3, 0.3) km s$^{-1}$, which is similar to that of 
Octans-Near, but apparently different in the W component.  However, given 
the variety of entries in VizieR for the proper motion of the binary and 
the radial velocity of the primary, the formal uncertainties on UVW quoted 
in the previous sentence are considerably too optimistic.

\subsection{Probable non-members}

We consider the four stars in Section C of Tables 1 and 2 to be 
non-members because of their apparent relatively old ages combined with 
UVW in poor agreement with that of the Octans-Near stars.

HIP 77810:  The Li EW = 103$\pm$5 m\AA\ is from Guillout et al (2009).  

HIP 102218:   The star is a rapid rotator (Table 2).

HIP 104526:  A 4" binary (G5 + G7) with respective Li EW 113 and 99 m\AA\ 
(Torres et al 2006).

HIP 108912:  Li EW = 110 m\AA\ (Cutispoto et al 2002; Torres et al 2006).  
The lithium EW is consistent with age $\sim$200 Myr.

\noappendix

\section*{References}
\begin{harvard}
\item[Anosova, J. \& Orlov, V. 1991, A\&A 252, 123]
\item[Arriagada, P., 2011, ApJ 734, 70]
\item[Barenfeld, S., Bubar, E., Mamajek, E. \& Young, P. 2013, ApJ 766, 6]
\item[Barnes, S. 2007, ApJ 669, 1167]
\item[Barrado y Navascues, D. 1998, A\&A 339, 831]
\item[Booth, M., Kennedy, G., Sibthorpe, B. et al. 2013, MNRAS 428, 1263]
\item[Caballero, J. 2010, A\&A 514, 98]
\item[Chen C., Patten B., Werner, M. et al. 2005, ApJ 634, 1372]
\item[Cutispoto, G., Pastori, L., Pasquini, L. et al. 2002, A\&A 384, 491]
\item[da Silva, L., Torres, C., de La Reza, R., et al. 2009, A\&A 508, 833]
\item[Dopita, M., Hart, J, McGregor, P., Oates, P \& Jones, D. 2007, Ap\&SS, 310, 255]
\item[Eggen, O. 1965, AR\&A 3, 235]
\item[Eisenbeiss, T., Ammler-von Eiff, M., Roell, T. et al 2013, A\&A 556, A53]
\item[Gerbaldi, M., Faraggiana, R. \& Balin, N. 2001, A\&A 379, 162]
\item[Glebocki, R. \& Gnacincki, P., 2005, "Proceedings of the 13th Cambridge Workshop on Cool] \indent Stars,
Stellar Systems and the Sun", ESA SP-560, ed. F. Favata et al., p. 571
\item[Gondcharov, G. 2006, Astro. Lett. 32, 759]
\item[Gray, R. O. \& Corbally, C. J. 2002, AJ 989, 2002]
\item[Gray, R. O., Corbally, C. J., Garrison, R. F., McFadden, M. T., Robinson, P. E., 2003, AJ 126, 2048]
\item[Gray, R. O., Corbally, C. J., Garrison, R. F. et al., 2006, AJ 132, 161]
\item[Guillout, P., Klutsch, A., Frasca, A. et al. 2009, A\&A 504, 829]
\item[Hauschildt, P., Allard, F., \& Baron, E., 1999, ApJ, 512, 377]
\item[Henry, T., Soderblom, D., Donahue, R., Baliunas, S., 1996, AJ 111,439.]
\item[Herrero, E., Ribas, I., Jordi, C., Guinan, E. F., Engle, S. G., 2012, A\&A, 537, 147]
\item[Isaacson, H., Fischer, D., 2010, ApJ 725, 875]
\item[Jura, M., Malkan, M., White, R. et al. 1998, ApJ 505, 897]
\item[Kenyon, S. \& Hartmann, L. 1995, ApJS 101, 117]
\item[King, J., Villareal, A., Soderblom, D., Gulliver, A. \& Adelman, S. 2003, AJ 125, 1980]
\item[Lindroos, K. 1985, A\&AS 60, 183]
\item[Lowrance, P., Schneider, G., Kirkpatrick, J. D. et al 2000, ApJ 541, 390]
\item[Maldonado, J., Mart$\acute{i}$nez-Arn$\acute{a}$iz, R., Eiroa, C., Montes, D. \& Montesinos, B. 2010, A\&A 521, 12]
\item[Mamajek, E. 2012, ApJ 754, L20]
\item[Mamajek, E. \& Hillenbrand, L. 2008, ApJ 687, 1264]
\item[Marois, C., Macintosh, B., Barman, T. et al 2008, Science 322, 1348]
\item[Martin, E., Magazzu, A. \& Rebolo, R. 1992, A\&A 257, 186]
\item[Monnier, J., Che, X., Zhao, M. et al 2012, 761, L3]
\item[Montes, D., L$\acute{o}$pez-Santiago, J., G$\acute{a}$lvez, M. et al. 2001, MNRAS 328, 45]
\item[Montesinos, B., Eiroa, C., Mora, A. \& Merin, B. 2009 A\&A 495, 901]
\item[Murgas, F., Jenkins, J. S., Rojo, P., Jones, H. R. A., Pinfield, D. J., 2013, A\&A, 552, 27]
\item[Nielsen, E. L., Liu, M. C., Wahhaj, Z., et al., 2013, arXiv:1306.1233v1, ApJ in press]
\item[Pace, G., 2013, A\&A, 551, 8]
\item[Pallavincini, R., Pasquini, L. \& Randich, S. 1992, A\&A 261, 245]
\item[Pecaut, M. \& Mamajek, E. 2013, arXiv:1307.2657, ApJ in press]
\item[Rhee, J., Song, I., Zuckerman, B. \& McElwain, M., 2007, ApJ 660, 1556]
\item[Riedel, A., Murphy, S., Henry, T. et al. 2011, AJ 142, 104]
\item[Rocha-Pinto, H. J., Flynn, C., Scalo, J., et al., 2004, A\&A, 423, 517]
\item[Rodriguez, D., Zuckerman, B., Kastner, J. et al., 2013, ApJ 774, 101]
\item[Rosvick, J., Mermilliod, J.-C. \& Mayor, M. 1992, A\&A 255, 130]
\item[Schmitt, J., Fleming, T. \& Giampapa, M. 1995, ApJ 450, 392]
\item[Schroeder, C., Reiners, A., Schmitt, J. H. M. M., 2009, A\&A, 493, 1099]
\item[Schroeder, C. \& Schmitt, J. H. M. M., 2007, A\&A 475, 677]
\item[Shkolnik, E., Anglada-Escude, G., Liu, M. et al. 2012, ApJ, 758 56]
\item[Soderblom, D., Jones, B., Balachandran, S., et al 1993, AJ 106, 1059]
\item[Stern, R. A., Schmitt, J. H. M. M., Kahabka, P. T., 1995, ApJ, 448, 683]
\item[Strassmeier, K., Fekel, F., Bopp, B., Dempsey, R., Henry, G., 1990, ApJS, 72, 191]
\item[Su, K., Rieke, G., Stansberry, J. et al. 2006, ApJ 653, 675] 
\item[Tetzlaff, N., Neuhauser, R. \& Hohle M. 2011, MNRAS 410, 190]
\item[Torres, C., Quast, G., Da Silva, L. et al 2006, A\&A 460, 695]
\item[Torres, C., Quast, G., Melo, C. \& Sterzik, M. 2008, 
"Handbook of Star Forming Regions,] \indent Volume II: The Southern Sky" 
ASP Monograph Publ., Vol. 5., ed. B. Reipurth, p.757
\item[van Leeuwen, F. 2007, A\&A 474, 653]
\item[Viana Almeida, P., Santos, N.,  Melo, C. et al. 2009, A\&A 501, 965]
\item[Vican, L. \& Schneider, A., 2013, submitted to ApJ]
\item[Vigan, A., Patience, J., Marois, C. et al. 2012 A\&A 544, 9]
\item[Vogt, S., 1987, PASP, 99, 621]
\item[White, R., Gabor, J. \& Hillenbrand, L. 2007, AJ 133, 2524] 
\item[Wright, J. T., Marcy, G. W., Butler, R., \& Vogt, S. 2004, ApJS 152, 261]
\item[Wright, E., Eisenhardt, P., Mainzer, A., et al., 2010, AJ, 140, 1868]
\item[Zuckerman, B. 2001, ARA\&A 39, 549]
\item[Zuckerman, B., Bessell, M. S., Song, I. \& Kim, S. 2006, ApJ 649, L115]
\item[Zuckerman, B., Rhee, J., Song, I., \& Bessell, M. 2011, ApJ 732, 61]
\item[Zuckerman, B. \& Song, I. 2004, ARA\&A 42, 685]
\item[Zuckerman, B. \& Song, I. 2012, ApJ 758, 77]
\item[Zuckerman, B., Song, I. \& Bessell, M. 2004, ApJ 613, L65]
\item[Zuckerman, B. \& Webb, R. 2000, ApJ 535, 959]

\end{harvard}

\clearpage

\begin{landscape}
\begin{deluxetable}{lccccccccccc}
\tablecolumns{12}
\tablewidth{0pt}
\setlength{\tabcolsep}{0.05in}
\tabletypesize{\scriptsize}
\tablecaption{Suggested Octans-Near Association Members} 
\tablehead{HIP& HD& R.A.& Decl.& Spec.& V& B-V& Dist.& Rad. Vel.& U,V,W (uncertainty)& Age$^{1}$ & Age$_{X-ray}$ 
\\
& (other)  & (h/m)& (deg)& Type& (mag)&  & (pc)& (km s$^{-1}$)& (km s$^{-1}$)&(Myr)&  (Myr)}  
\startdata
\multicolumn{12}{c}{A. Probable members of the Octans Near Association} \\
\hline
813&560& 00 10&11& B9+K1&5.5&-0.07&94& 13.4$\pm$0.8& -14.9, +2.2, -11.9 (0.4, 0.5, 0.6)  &30&\\
10670&14055& 02 17&33& A1&4&0.019&34& 9.9$\pm$3.1&-11.3, -3.9, -9.0 (2.2, 1.7, 1.3) &100&\\
14007&18809& 03 00&-37& G3+K5&8.5&0.677&46& 16.1$\pm$0.3& -10.5, -5.5, -12.9 (0.3, 0.2, 0.3)  &100&100\\
16095&21379& 03 27&12& A0&6.3&-0.025&88& 16.6$\pm$3.1& -14.7, -3.3, -8.8 (2.5, 0.5, 1.8) &$\leq$100&\\
17338&23208& 03 42&-20& G8&9.2&0.78&50& 17.4$\pm$0.3& -13.5, -2.7, -12.0 (0.4, 0.3, 0.3) &30&30\\
19496&283420& 04 10&25& F8&9.1&0.44&98& 15.3$\pm$1.1& -13.7, -4.6, -10.3 (1.0, 0.9, 0.8) &30&young?\\
19990&27045& 04 17&20& A3m&4.9&0.259&29& 15.0$\pm$0.6& -10.9, -1.7, -14.2 (0.6, 0.1, 0.2) &100&\\
22192&30422& 04 46&-28& A3&6.2&0.189&56& 14.4$\pm$1.0& -11.2, -5.4, -8.7 (0.5, 0.5, 0.6)  & $\leq$100 &\\
36624&59507& 07 31&38& A2&6.5&0.077&81& 7.0$\pm4.4$& -13.4, -4.7, -12.3 (4.0, 0.3, 2.0) &$\leq$100&\\
73765&133725& 15 04&38& F8&7.6&0.503&49& -13.8$\pm$3.6& -11.5, -3.8, -11.0 (0.9, 1.6, 3.1)  &100&300\\
97255&186704& 19 45&4& G0+M0&7&0.612&31& -10.1$\pm$0.4& -12.2, -4.4, -8.5 
(0.3, 0.3 0.3) &100 &50\\
102333&197157& 20 44&-51& A9&4.5&0.278&24& -1.6$\pm$0.8& -13.5, -4.4, -12.6 (0.6, 0.1, 0.5) &100&\\
105044&202398& 21 16&-36& F5&8.1&0.485&70& -0.1$\pm$0.4& -14.3, -1.1, -14.4 
(0.8, 0.2, 0.8)  & 100 &100\\
115527&220476& 23 24&-7& G5&7.6&0.682&30& 4.4$\pm$0.2& -17.0, -6.1, -10.7 (0.4, 0.2, 0.2) &100&100\\
\hline
\multicolumn{12}{c}{B. Possible members of the Octans Near Association} \\
\hline
29873&42994& 06 17&57& F8&7.1&0.523&55& 10.6$\pm$0.3& -17.7, -3.1, -11.6 (0.7, 0.6, 1.2)  &200&300\\
44526&77825& 09 04&-15& K2&8.8&0.96&28& 2.8$\pm$0.8& -8.4, -4.3, -12.0 (0.4, 0.7, 0.4) &150&300\\
66704&119124& 13 40&50& F8+K7 &6.4&0.537&25& -12.2$\pm$0.3& -14.7, -9.4, -10.8 (0.2, 0.1, 0.3)&200&300\\
104864&202116& 21 14&-22& G2&8.4&0.61&51& -2.0$\pm$0.2& -18.1, -3.3, -16.9 (0.8, 0.2, 0.9) &200?&70\\
 & GJ 4185& 21 16& +29& M3.3 + M3.3& 12.7& 1.3& 19.4& -2.0$\pm$0.3&  -16.2, -1.2, -10.7 (1.2, 0.5, 1.2) & 150$^{2}$ & \\
110778&212698& 22 26&-16& G2 + G3&6.3&0.62&20& +1:& -18, -7, -14 &150&300\\
116132& & 23 31&19& M3.5 + M4.5&10.3&1.4&6.2& 1.5$\pm$0.4& -13.6, -6.0, 
-7.0 (0.2, 0.3, 0.3) &$<$100$^{2}$ & $\leq$100\\
\hline
\multicolumn{12}{c}{C. Probable non-members}\\
\hline
77810&142229& 15 53&4& G5&8.1&0.63&41& -21.1$\pm$0.2& -18.3, -5.5, -10.1 (0.3, 0.2, 0.3)  &200&300\\
102218&197211& 20 42&-10& F5&7.7&0.48&67& -5.2$\pm$2.4& -9.8, -4.1, -7.5 (1.7, 1.2, 1.4) &$>$200?&600\\
104526&201247& 21 10&-54& G5 + G7&7.1&0.69&35& -3.1$\pm$0.3& -17.2, -3.0, -12.4 (0.5, 0.2, 0.5) &200&300\\
108912&209234& 22 03&-60& G2&7.9&0.62&41& -2.1$\pm$0.2& -17.6, -2.1, -10.7 (0.6, 0.2, 0.5) &200&300\\\\

\enddata

\tablecomments{The UVW of the Octans Association defined by Torres et al 
(2008) is
-14.5$\pm$0.9, -3.6$\pm$1.6, -11.2$\pm$1.4 km s$^{-1}$.  UVW are given 
with respect to the Sun with U positive toward the Galactic Center, V 
positive in the direction of Galactic rotation, and W positive toward the 
North Galactic pole.  The UVW of the 14 systems listed in Section A of Table 1 is 
-13.0 $\pm$1.9, -3.5$\pm$2.2, -11.2$\pm$2.0 km s$^{-1}$; if HIP 813 is not 
included then the mean V for the other 13 systems is -4.0$\pm$1.5 (see Section 3).
CD-58 860, previously suggested by Torres et al to be a 
member of the Octans Association, is also a plausible member of Octans-Near 
at a photometric distance from Earth of $\sim$90 pc (see Section 4) and
position 4h11m and -58d. \\
$^{1}$ The ages listed in the penultimate column for B- and A-type stars 
are from location on an M$_V$ vs B-V color-magnitude diagram and from 
excess infrared emission (Section 4.1).  Ages for spectral type stars F, G, 
and K are based on chromospheric and coronal activity and lithium 6708$\AA$ 
EW calibrated with Figure 3 and with Figure 3 of Zuckerman \& Song (2004).\\
$^{2}$ The ages listed in the penultimate column for GJ 8185 and HIP 116132 
are based on placement on color magnitude diagrams (Shkolnik 
et al 2012; Rodriguez et al 2013; Zuckerman \& Song 2004; Riedel et al 2011).}
\end{deluxetable}
\end{landscape}

\clearpage

\begin{deluxetable}{cccccccccc}
\tablewidth{0pt}
\tabletypesize{\scriptsize}
\tablecolumns{8}
\tablecaption{Age Data for Suggested Octans-Near Members}
\tablehead{
\colhead{HIP}&
\colhead{vsini (lit)}&
\colhead{vsini (calc)} &
\colhead{Li EW}&
\colhead{Li EW ref}&
\colhead{f$_{x}$}&
\colhead{logR'$_{HK}$}&
\colhead{logR'$_{HK}$ ref}&\\
\colhead{}&
\colhead{(km s$^{-1}$)}&
\colhead{(km s$^{-1}$)}&
\colhead{(m$\AA$)}&
\colhead{}&
\colhead{}&
\colhead{}&
\colhead{}}
\startdata
\multicolumn{8}{c}{A. Probable members of the Octans Near Association}\\
\hline
813&240&&290$^{1}$ &P92, M92&&&\\
10670&242&&&&&&\\
14007&9.5&&132&C02, T06&-4.35&-4.33&a,b,h,i\\
16095&70&&&&&&\\
17338&9&11.1&258&this work&-3.25&-4.169&f,j\\
19496&26.6&26.9&142&this work&-4.62&&\\
19990&70.1&&&&&&\\22192&124&&&&&&\\
36624&118&&&&&&\\
73765&29&&83&this work&-4.58&&\\
97255&13.5&14.9&120&this work&-3.96&-4.35&d,i\\
97255B&&44.3&$<$15&this work&&&\\
102333&150&&&&&&\\
105044&15& & 64 &  Appendix A1 &-4.48&&\\
115527&4.7&2&135&this work&-4.2&-4.39&e,g,h,i,j\\ \hline 
\multicolumn{8}{c}{B. Possible members of the Octans Near Association}\\\hline
29873&12&9.7&77& this work&-4.42&&\\
44526&9&7&55&this work&-4.07&-4.409&h\\
66704&10&10.2&106&this work&-4.46&-4.374&b,c,i,j\\
66704B&&  &$<$30&this work&&&\\
104864&8.1&&85&this work&-3.97&&\\
GJ4185&&&&&$\sim$-2.8&&&\\
110778&12.8&&122&this work&-4.46&-4.479&h\\
110778B&8.5&6.1&143& this work&&&\\
116132&10&7.8&&&-2.7&&\\\hline

\multicolumn{8}{c}{C. Probable non-members}\\\hline
77810&5&&103&G09&-4.29&-4.353&d,i,j,k,l \\102218&&37.4 &36&this work&-4.96&&\\
104526&4&&113+99& T06 &-4.29&-4.4&a,i\\108912&11&&110&C02, T06& -4.42 &-4.42&b,h,i
\enddata
\tablecomments{logR'$_{HK}$ references: a=Henry+1996, b=Rocha-Pinto+1998, c=Gray+2003, d=Wright+2004, e=Gray+2006, f=White+2007, g=Arriagada+2011, h=Herrero+2012, i=Murgas+2013, j=Pace2013, k=Isaacson+2010, l=Schroeder+2009\\
$^{1}$The listed lithium EW for HIP 813 is for the K1 secondary. \\
Li EW references: P92= Pallavicini et al. 1992, M92 = Martin et al. 1992, C02 = Cutispoto et al. 2002, T06 = Torres et al. 2006, G09 = 
Guillout et al. 2009\\
vsini(calc) are from our Hamilton spectra, calculated using absorption features near 6430$\AA$.\\
vsini(lit) are taken from the literature, largely from Glebocki \& Gnacincki 2005. \\
Where "this work" appears in the fifth column, Li EW are from the Hamilton spectra. \\
f$_{x}$ = L$_{x}$/L$_{bol}$ where L$_{x}$ is from the ROSAT All-Sky Survey.\\
Errors in the listed values of vsini(calc) and Li EW are given in the last two paragraphs of section 2.}
\end{deluxetable}

\clearpage
\begin{deluxetable}{ccccc}
\tablewidth{0pt}
\tabletypesize{\scriptsize}
\tablecolumns{8}
\tablecaption{Spectroscopic Settings}
\tablehead{
\colhead{Spectrometer}&
\colhead{UT Date}&
\colhead{Slit Width (")} &
\colhead{Resolution (R)}&
\colhead{Useful Wavelength Range($\AA$)}}
\startdata
Kast Red & Oct 31 & 1.0 & 2,500 & 5800-7200\\
Kast Blue & Oct 31 & 1.0 & 1,100 & 3300-5520\\
Hamilton & Nov 25 & 2.5 & 45,000 & 3500-7500\\
Hamilton & Nov 26 & 2.5 & 45,000 & 3500-7500\\
Hamilton & Nov 27 & 2.5 & 45,000 & 3500-7500\\
Hamilton & Feb 22 & 2.5 & 45,000 & 3500-7500\\
Hamilton & Mar 24 & 2.5 & 45,000 & 3500-7500\\
Kast Red & May 22 & 1.0 & 2,500 & 5800-7200\\
Kast Blue & May 22 & 1.0 & 1,100 & 3300-5520\\
\enddata
\tablecomments{For Kast observations, the 1200/5000 grating was used on the red side, while the 600/4310 grism was used on the blue side.}
\end{deluxetable}

\clearpage
\begin{landscape}
\begin{figure}
\includegraphics[width=200mm]{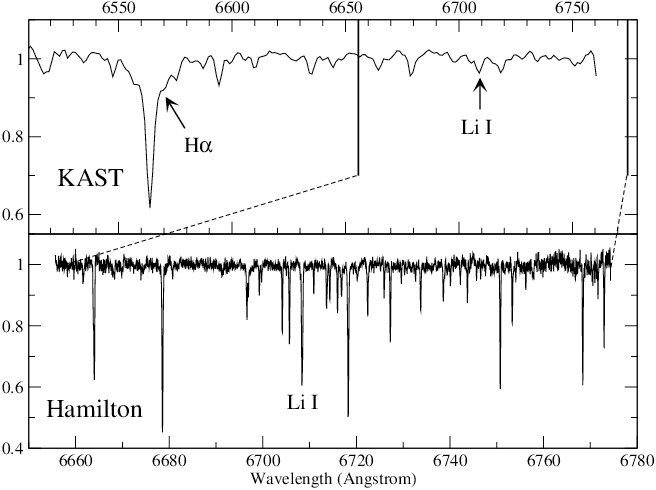}
\caption{\label{figure 1} Representative Kast and Hamilton spectra showing 
the Li 6708 \AA\ absorption line in the Octans-Near star HIP 115527. The 
abscissa is in the heliocentric rest frame.}
\end{figure}   
\end{landscape}

\clearpage
\begin{landscape}
\begin{figure}
\includegraphics[width=250mm]{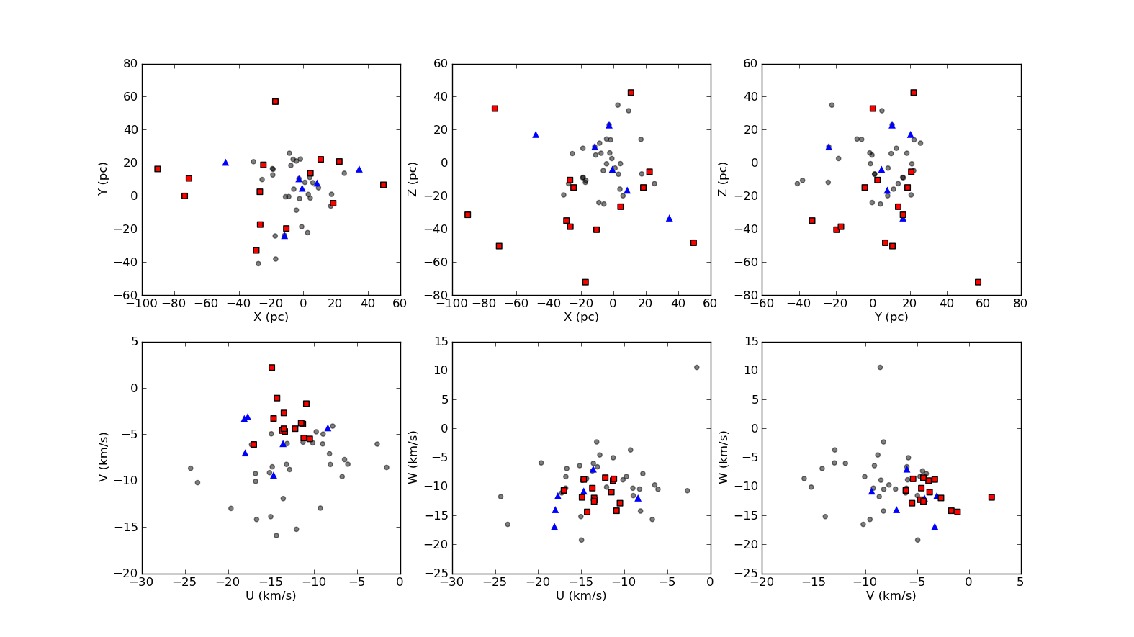}
\caption{\label{figure 2}\emph{Top:} The XYZ spatial distributions of 
proposed Octans-Near members. Red squares are probable members, blue 
triangles are possible members, and grey circles are proposed members of 
the Castor moving group (from Montes et al 2001; see Section 4.2). 
\emph{Bottom:} The UVW Galactic space motions of proposed Octans-Near and Castor 
members.}
\end{figure}
\end{landscape}

\clearpage
\begin{landscape}
\begin{figure}
\includegraphics[width=250mm]{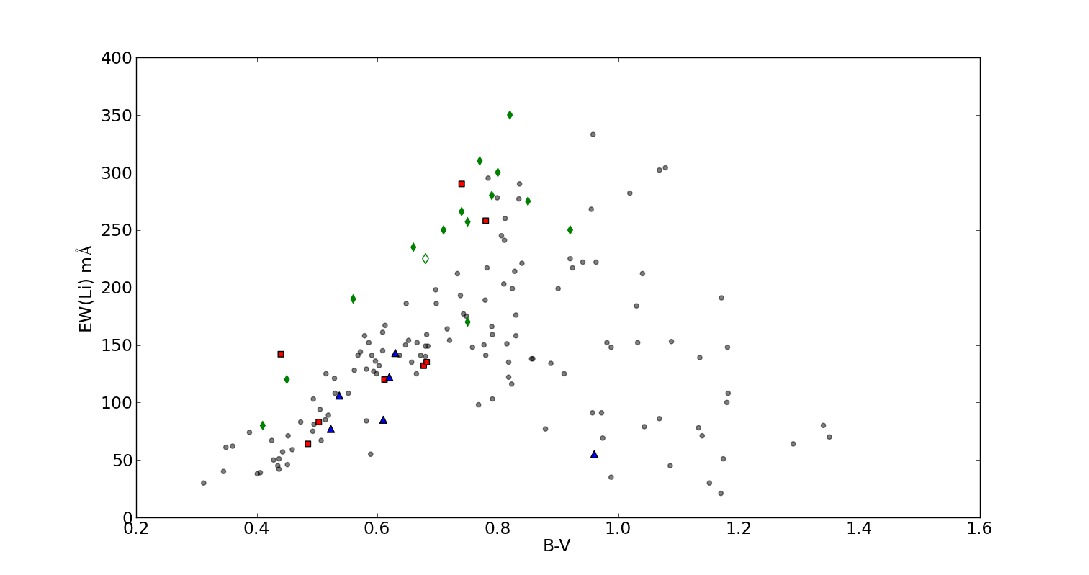}
\caption{\label{figure 3} Equivalent width of the lithium 6708 \AA\ line 
for proposed Octans-Near members, Octans Association members (da Silva 
et al 2009), and Pleiades members (Soderblom et al 1993). The red 
squares are probable Octans-Near members from Table 1A, the blue triangles 
are possible members in Table 1B, the filled green diamonds are Octans 
Association members, and the grey circles are Pleiads.  The red square 
with B-V = 0.74 and Li EW = 290 m\AA\ is HIP 813B.  The one unfilled 
green diamond is CD-58 860 which is a plausible member of both Octans-Near 
and the Octans Association.}
\end{figure}
\end{landscape}

\clearpage
\begin{landscape}
\begin{figure}
\includegraphics[width=200mm]{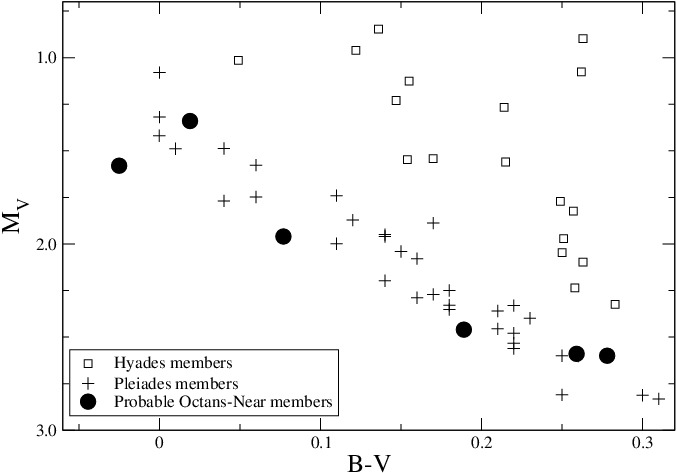}
\caption{\label{figure 4} Color magnitude diagram for the six A-type stars in 
Table 1A.  As noted in Appendix A1, to compensate for its $\lambda$ Boo 
nature, HIP 22192 should be plotted slightly redward of its conventional 
B-V of 0.19 (given in Table 1A).  B-type star HIP 813A is too blue to be plotted 
in this figure, but it falls very near to a Pleiades star plotted on Figure 5 of 
Zuckerman (2001). }      
\end{figure}   
\end{landscape}

\clearpage
\begin{landscape}
\begin{figure}
\includegraphics[width=250mm]{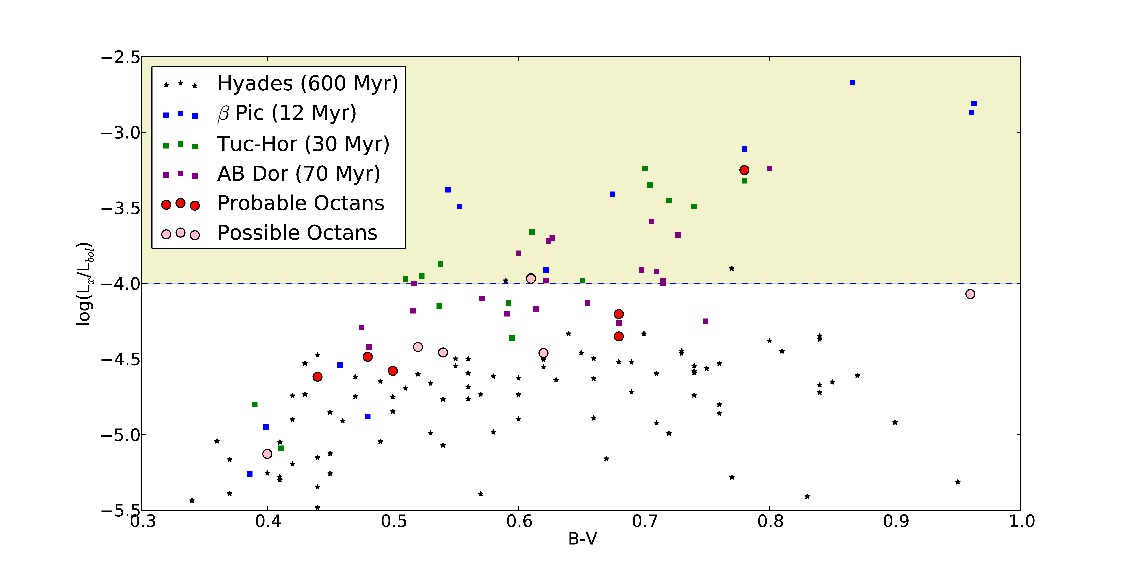}
\caption{\label{figure 5}Empirical relation between X-ray activity and age as a 
function of B-V color. This figure was used to estimate the ages of stars 
(Age$_{X-ray}$) listed in Table 1. Members of young nearby moving groups are from 
Torres et al. (2008) and Hyades members are from Stern et al. (1995). The 
horizontal line represents the point at which the Mamajek \& Hillenbrand (2008) 
relationship between X-ray luminosity and the chromospheric index (logR'$_{HK}$) 
ceases to apply.}
\end{figure}
\end{landscape}


\clearpage
\begin{figure}
\center
\includegraphics[width=150mm]{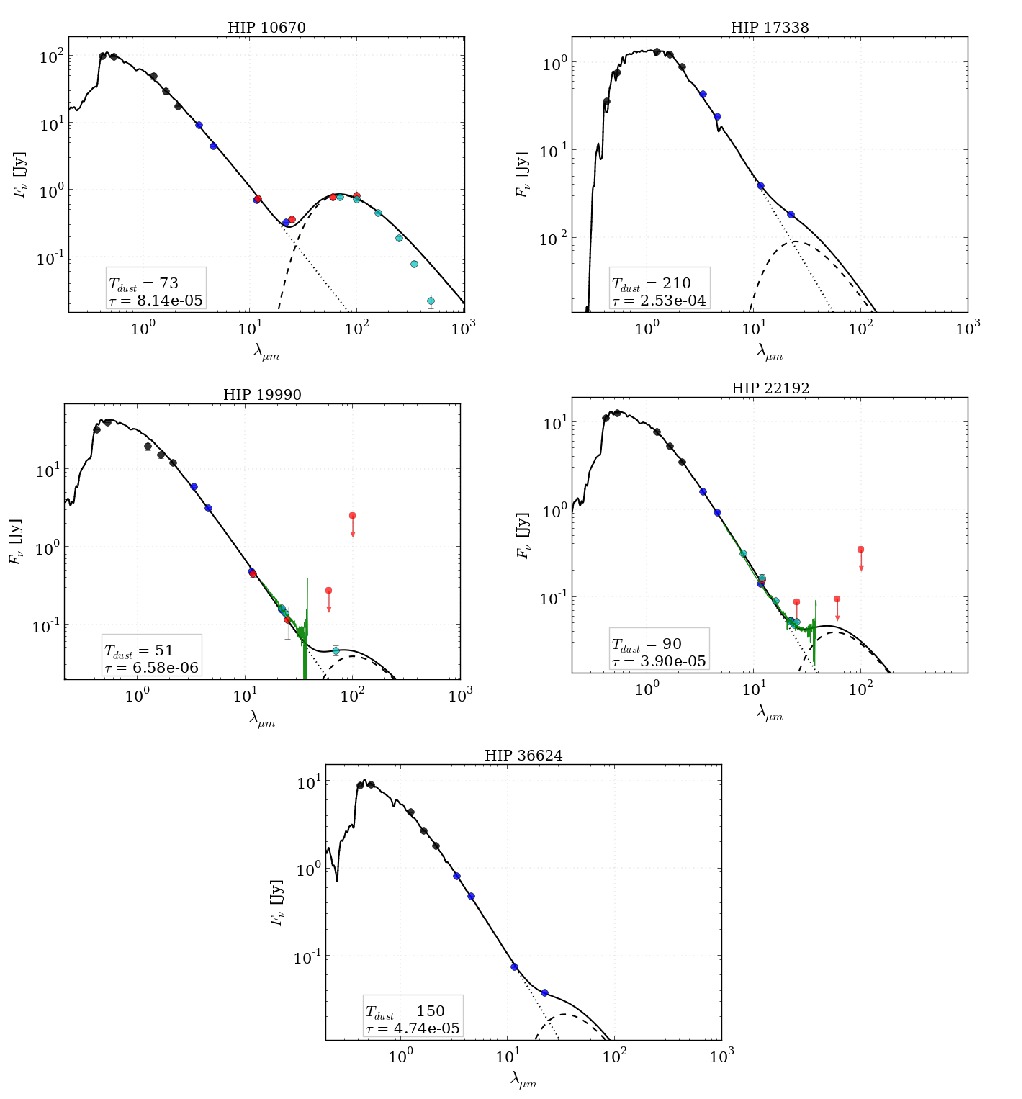}
\caption{Spectral energy distributions for five candidate 
Octans-Near members with definite IR excesses. The WISE data products were 
checked for contamination by a background source or IR cirrus. B and V 
magnitudes come from the Tycho catalog and the J, H, and K data  
are from the 2MASS catalog; these 5 data points are shown in 
black. Additional data from IRAS and Spitzer IRAC and MIPS were added 
when available. IRAS data are shown in red, while IRAC and MIPS data 
are shown in light blue. For two stars, IRS spectra from Spitzer were 
available (green line). For HIP 10670 Herschel data (the 3 longest 
wavelength points in blue) were available from the DEBRIS survey (Booth et 
al 2013). \label{figure6}}
\end{figure}

\clearpage
\begin{figure}
\includegraphics[width=125mm]{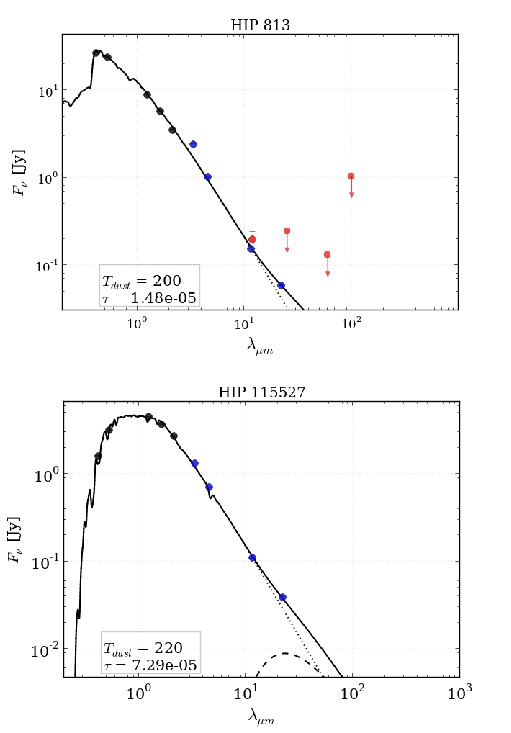}
\caption{\label{figure 7}Spectral energy distributions for two 
proposed Octans-Near members with possible infrared excess in WISE Band 4 
(22 $\mu$m). The WISE data products were checked for contamination by a 
background source or IR cirrus. The excess around HIP 115527 is only at the 
3.5$\sigma$ level, and is therefore considered to be uncertain. Although 
HIP 813 has an apparent $\sim$7$\sigma$ W4 excess, it also appears to have 
an excess at W1 (3.4 $\mu$m). While WISE is known to have a problem at W2 
where fluxes above 1 Jy cannot be trusted, W1 is not known to suffer in a similar
way. Thus, the origin of the high data point at W1 is unclear to the 
authors.}
\end{figure}

\end{document}